\documentclass[12pt]{article}
\usepackage{graphics}
\input epsf

\newcommand{\be}{\begin{eqnarray}}
\newcommand{\ee}{\end{eqnarray}}

\begin{document}
\title{
\begin{flushright}
{\large UAHEP051}
\end{flushright}
\vskip 1cm
Gravitational collapse of a susy star}
\author{L. Clavelli\footnote{lclavell@bama.ua.edu}\\
Department of Physics and Astronomy\\
University of Alabama\\
Tuscaloosa AL 35487\\ }
%\today
\date{February 1, 2006}%\\ \today}
\maketitle
\begin{abstract}
The evidence for a positive vacuum energy in our universe suggests that
we might be living in a false vacuum destined to ultimately decay to
a true vacuum free of dark energy.  At present the simplest
example of such a universe is one that is exactly supersymmetric (susy).
It is expected that the nucleation rate of critically sized susy bubbles
will be enhanced in regions of high density such as in degenerate stars.
The consequent release of energy stored in Pauli towers provides a
possible model for gamma ray bursts. Whether or not all or any of the
currently observed bursts are due to this mechanism, it is important
to define the signatures of this susy phase transition. After such a burst,
due to the lifting of degeneracy pressure, the star would
be expected to collapse into a black hole even though its mass is below
the Chandrasekhar limit.  Previous studies have treated the star as
fully releasing its stored energy before the collapse.  In this article
we make an initial investigation of the effects of the collapse during
the gamma ray emission.
\end{abstract}

%\pacs{PACS numbers: 11.30.Pb, 12.60.Jv, 13.85.-t}
%\ccode{PACS Nos.: 11.30.Pb, 12.60.J, 13.85.-t}
\renewcommand{\theequation}{\thesection.\arabic{equation}}
\renewcommand{\thesection}{\arabic{section}}
\section{\bf Introduction}
\setcounter{equation}{0}

Present indications are that we live in a broken-susy universe with 
a positive vacuum energy density
\be
    \epsilon = 3560 {\displaystyle MeV/m}^3
\label{vacenergy}
\ee
leading to an acceleration in the expansion of the universe.  This has
given new fuel to the idea that we live in a false vacuum and that the
universe will ultimately make a transition to a lower energy vacuum.
If, as suggested by string theory, there is an exactly supersymmetric,
zero vacuum energy universe that is dynamically connected to the broken
susy universe, the final phase of the universe could be exactly 
supersymmetric.  The theory of vacuum decay was pioneered some decades ago by
Coleman \cite{Coleman}.  In this theory, bubbles of true vacuum are
continually being
nucleated in the false vacuum.  Most of these are quite small and are
immediately quenched.
However, when one appears with radius greater than some critical radius
\be
     R_c = \frac{3S}{\epsilon}
\label{criticalradius}
\ee
it will rapidly grow to take over the universe.  Here, $S$ is the
surface tension of the bubble assumed to be independent of bubble size.
The probability per unit time per unit volume to produce a bubble of
radius $R_c$ or greater and, therefore, to effect a phase transition to
the true vacuum is given \cite{Coleman} in the form
\be
     \frac{d^2P}{dt d^3 r} = A e^{-B(vac)}
\label{AemB}
\ee
where, assuming a thin wall between the phases,
\be
     B(vac) = \frac{27 \pi^2 S^4}{2 \epsilon^3 } \quad.
\label{Bvac}
\ee

A first look at the environment of a susy universe has been reported in
\cite{future} but, for sufficiently large $S$ and small $A$, the transition
is not likely to take place in the near future.
In much of this article we use the solar mass, $M_\circ = 1.2
\cdot 10^{60}$ MeV, and earth radius, $R_E = 6.38 \cdot 10^6$ m as
convenient units.  Factors
of $\hbar$ and $c$ are sometimes left implicit.
%We also use natural units $\hbar=c=1$.

\begin{figure}%[htbp]
\begin{center}
\epsfxsize= 4in %% 6.7in % actual
% 4.5in
\leavevmode
\epsfbox{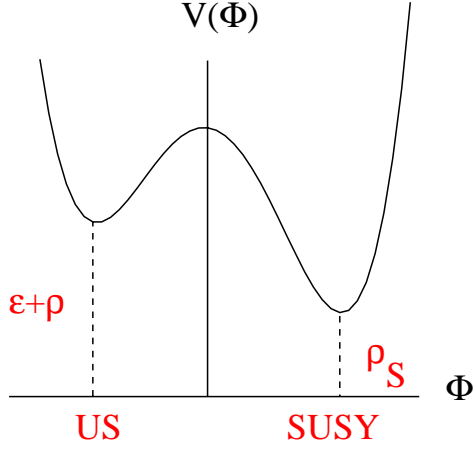}
\end{center}
\caption{The effective potential showing the false vacuum of broken susy and the
true vacuum of exact susy.}
\label{susywell}
\end{figure}

     Reasonable arguments \cite{Gorsky,Voloshin} have been given
that the transition rate could be enhanced in dense matter.
This enhancement can be implemented in a natural way \cite{CK} 
if the
above equations are modified in dense matter by replacing $\epsilon$
by the total energy advantage per unit volume of trading the broken
susy phase for the exact susy phase, i.e.
\be
      \epsilon \rightarrow \epsilon + \Delta \rho\\ \nonumber
      R_c \rightarrow \frac{3S}{\epsilon + \Delta \rho}
\label{mattereffect}
\ee
where $\Delta \rho$ is the difference in the ground state matter densities
between the broken susy phase and the exact susy phase as shown in fig.\  \ref{susywell}.

The difference $\Delta \rho$ is the fermionic
excitation energy density.  The parameter controlling the exponential
factor in the
transition rate would then be
\be
      B = \frac{27 \pi^2 S^4}{2 (\epsilon+\Delta\rho)^3 } \quad.
\label{Bmatter}
\ee
The value of $\Delta \rho$ in a white
dwarf star is calculated as follows.  In a degenerate electron gas of 
$N_e$ electrons in a volume $V$ the Fermi momentum is
\be
      p_F = \left ( \frac{3 \pi^2 N_e}{V} \right) ^{1/3} \hbar
\label{Fermimomentum}
\ee
with, assuming equal numbers of neutrons and protons,
\be
      N_e/V = \frac{\rho}{2 M_N} f_e
\label{Ne}
\ee
Here $M_N$ is the nucleon mass and $f_e$ is the electron to proton ratio.
Before the phase transition
$f_e$ is equal to unity but afterwards it decreases as electron pairs
convert to selectron pairs.  At any stage in the conversion process,
\be
       e^{-}e^{-} \rightarrow {\tilde e}^{-}{\tilde e}^{-}\quad ,
\label{sigma}
\ee
one would have, by charge conservation, $N_P = N_e + N_{\tilde e}$, so
\be
     f_e = \frac{N_e}{N_e+N_{\tilde e}}
\ee
In previous studies \cite{growth}, this conversion process reducing $f_e$ to
essentially zero was assumed for simplicity to take place at fixed stellar
radius.  In actuality, of course, as the Fermi degeneracy is lifted the
star will tend to shrink under the force of gravity.  Without attempting
at this time to treat the full time-dependent problem we give in this 
paper some attention to sequential stages in the gravitational collapse
of a susy star.  

The average electron kinetic energy in the degenerate gas of a white dwarf
star is
\be
      <E> -m = m \left( -1 + \frac{3}{8} {\sqrt{(1+z)}} (2+1/z)
       - \frac{3}{8z^{3/2}} \ln \frac{\sqrt{z} -1 + \sqrt{1+z}}
         {\sqrt{z}+1-\sqrt{1+z}}\right) \quad.
\label{exact}
\ee
where
\be
      z = \frac{p_F^2}{m^2} \quad .
\ee
In the limit of high Fermi momentum relative to the electron mass, this is
\be
      <E> -m = \frac{3 p_F}{4} \quad.
\label{himomlimit}
\ee
while for Fermi momentum low compared to the electron mass,
\be 
       <E> - m = \frac{3}{10} m z \quad .
\ee
The kinetic energy density, which is equal to the difference in ground state
energy densities between the broken susy state and the exact susy state, is then
\be
      \Delta \rho(r) = (<E>-m) \frac{\rho}{2 M_N} f_e
\label{delrho}
\ee
We neglect for now contributions from
nuclear excitation energies.  In \cite{growth}, we used for 
simplicity the high Fermi momentum limit \ref{himomlimit}.  In our current
work the exact expression \ref{exact} is used.

In the broken susy phase ($f_e=1$), a white dwarf star is supported by a balance between
the outward electron degeneracy pressure gradient and the inward gravitational
pressure gradient.  The mass and radius of a white dwarf star are then 
each determined by the density at the center leading to a unique mass-radius
relation.  These relationships, originally calculated by Chandrasekhar
\cite{Chandra} are shown in \cite{growth} at zero temperature.  
In section II we calculate the 
corresponding relations for several cases of $f_e$ below unity but assumed
uniform over the volume of the star.  In the case of uniform $f_e$, the
star would collapse at constant mass though these stages of decreasing $f_e$ 
(assuming the
radiation is a negligible fraction of its rest energy during the process).
As the radius of the star decreases, the central density and, therefore,
the energy stored in the Pauli tower
of electrons increases even though the fraction of Fermionic electrons is
decreasing.  Thus the collapse provides a natural pumping mechanism from
gravitational energy into the stored Pauli energy.
It is found that the central density increases so rapidly during this
process that a central core of the star decouples and becomes a black hole 
while $f_e$ is still relatively high.

This calculation, however, is obviously oversimplified since the 
critical radius of 
the susy bubble is expected to be small compared to the radius of the star.
$f_e$ would drop rapidly to near zero within the bubble which would then only
slowly expand to engulf the star.  It also neglects the radiation pressure
due to the Pauli energy released within the bubble.

In section III we slightly refine the toy model to take into account
in an adiabatic and very approximate way
a non-uniformity of $f_e$ and the effect of nuclear energy release.
We assume that the susy bubble 
begins at the center of the star with a critical radius and then proceeds 
through stages of
greater radius.  Inside the bubble the degeneracy pressure is set to zero
but there is a balance between the gravitational pressure gradient and the
radiation pressure gradient due to the nuclear energy release.  Outside the
bubble wall there is the usual balance between the degeneracy pressure 
gradient and the gravitational pressure gradient.  Photons of energy less
than the Fermi energy of the electron gas are assumed to pass freely out
of the star due to the Landau-Pomeranchuk effect \cite{Landau} 
but most of the radiation
is trapped in the regions of high density and only slowly cools.  We 
neglect this cooling within the bubble which, of course, would ultimately
lead again to a black hole collapse due to the absence of Fermion degeneracy.
The sudden dumping of nuclear energy near the stellar center might also
set up density waves throughout the star which could affect the growth
of the bubble.
The complete time dependence of this process would require a much more
elaborate Monte-Carlo calculation than we can attempt at present.

In both of the simplified models treated here, the energy released in the
collapse could be significantly greater than that calculated
in \cite{growth} although this depends to some extent on how much of the
energy release is trapped within the developing black hole.  In both cases
the mass of the resulting black hole would still be significantly below the
Chandrasekhar limit which was one of the prime predictions of
the original susy star model.

Section IV is devoted to a summary of our current level of
understanding of stellar behavior following a phase transition
to exact supersymmetry. 

\section{\bf Stages of uniformly decreasing $N_e/N_p$}
\setcounter{equation}{0}

    Following a susy phase transition, the degeneracy pressure in
a dense star will decrease as electron pairs convert to scalar
electron (selectron) pairs via photino exchange.  In this section
we neglect radiation pressure and treat, as a toy model, the case
of a uniform ratio of electron to selectron numbers throughout the
star.  Such a model might be more realistic if the surface tension
of eq.\,\ref{mattereffect} was such as to make the critical
radius comparable to the radius of the dense star. 

These approximations will be somewhat relaxed in the subsequent section.
At present we seek to determine the density profile of the star
as a function of $f_e$ at fixed stellar mass.  We follow the
Chandrasekhar calculation but allow for a sequence of
decreasing $f_e$ ratios.

    The gravitational pressure gradient is

\be
    \frac{dP_G}{dr} = - \rho(r) G \frac{M(r)}{r^2}
\label{gravgrad}
\ee
where Newton's constant, in convenient units of earth radius and
solar mass, is
\be
   G = 2.34\cdot 10^{-4} R_E c^2 M_0^{-1}\quad.
\ee
At zero temperature, a stable density profile is defined
by a balance between this gradient and the gradient of degeneracy
presssure.  The degeneracy pressure is
\be
    P_d = a f(x)
\ee
where
\be
    x = \frac{P_F}{mc} = b (\rho f_e)^{1/3}
\ee
and
\be
    a =& \frac{m^4 c^5}{3 \pi^2 \hbar^3} = 0.0165 R_E^{-3} M_0 c^2 \qquad &\\ \nonumber
    b =& \frac{2\pi \hbar}{mc}(\frac{3}{8\pi m_N \mu_e})^{1/3}
    = 1.6 R_E M_0^{-1/3} & \quad.
\ee
Here $\mu_e= A/Z$ which we take equal to $2$ as in a Carbon or
Oxygen star.  These formulae are straightforward generalizations to $f_e$
below unity from standard presentations as, for example, given
in \cite{Harwit}.  The function $f(x)$ is given by
\be
 f(x) = \frac{1}{8} \left ( x(2 x^2 -3)\sqrt{x^2+1} + 3 \sinh^{-1}(x) \right )
\ee
so that
\be
     \frac{dP_d}{dr} = \frac{ab}{3} (f_e \rho)^{-2/3} f^\prime(x) \frac{d(\rho f_e)}{dr} \quad.
\label{Pdgrad}
\ee
where
\be
     f^\prime(x) = \frac{x^4}{\sqrt{1+x^2}}\quad.
\ee
The balance of gravitational and degeneracy pressure requires that
\be
   f_e \frac{dw}{dr} = - \frac{G}{ab^3} \frac{M(r)}{r^2}
\label{wequation}
\ee
where
\be
    w = \sqrt{1+ b^2 (\rho f_e)^{2/3}}\quad .
\ee
Since
\be
    M(r) = 4\pi \int_{0}^{r} {r\,^\prime}^2 dr\,^\prime \rho(r\,^\prime)
\ee
The equilibrium density profile satisfies the second order
differential equation
\be
  \frac{1}{r^2} \frac{d}{dr} f_e r^2 \frac{dw}{dr} =
          - \frac{4\pi G}{ab^3} \rho(r)\quad .
\ee
We find it, however, more convenient to use eq.\ref{wequation}.
We begin by specifying some central density  $\rho(0)$ with
$M(0)=0$.  We then integrate out in steps of $dr=M_0\cdot 10^{-5}$ 
putting
\be
    M(r+dr) = M(r) + 4 \pi \rho(r) r^2 dr
\ee
and
\be
    w(r+dr) = w(r) + \frac{dw}{dr} dr \quad.
\ee
The gravitational energy is zero at the center and
is incremented according to
\be
    U(f_e, r+dr) = U(f_e, r) - 4\pi G r dr M(r) \rho(r)\quad .
\ee
The process terminates at the point at which the density drops
to zero and this defines the radius and mass of the star as well
as its total gravitational energy.

We consider a star with central density $\rho_0= 8.3 M_0 R_E^{-3}$.
Before the susy phase transition ($f_e=1$) such a star will be stable
at a
mass of $M = 1.09 M_0$ and a radius of $R(f_e=1) = 0.73 R_E$.
Its total gravitational potential energy is found to be
\be
   U(1,R) = - 1.6\cdot 10^{55} {\displaystyle{ergs}} \quad .
\ee
We then decrease $f_e$ by a small amount and repeat the process
increasing $\rho(0)$ so that the total mass remains the same.
These stages of stepwise decreasing $f_e$ are characterized by
a sequence of
decreasing radii $R(f_e)$, increasing central densities
$\rho_0(f_e)$, and increasingly negative gravitational energy.
Due to the increasing density, the local Fermi momenta as given 
by eqs.\ref{Fermimomentum}
and \ref{Ne} increase in this process even though
the electron fraction $f_e$ is decreasing.  Thus the collapse
provides a natural pumping mechanism from gravitational
energy into the energy of the electron cloud.  This energy is
released as electrons convert to selectrons which are not bound
by the Pauli principle.  The resulting photons of energy less
than the Fermi energy escape from the star with little absorption
due to the Landau-Pomeranchuk effect.  This mechanism for a fast
escape of photons from a star was first pointed out by Takahashi
et al. \cite{Takahashi}.

The Schwarzschild radius for a spherical object of mass $M(r)$ is
\be
     R_S(r) = \frac{2 G M(r)}{c^2} \quad.
\ee
If at any point in the integration,
\be
     r < R_S(r)
\ee
the stellar core decouples and collapses to a black hole.
This core collapse is analogous to that of the collapsar model
\cite{collapsar} of long gamma ray bursts.

The collapsar
model is, however, most obviously applicable to heavy stars whereas we
are dealing with a near solar mass star.  Attempts to describe short
bursts ($\tau < 2$ s) within the standard model are often based on the
hypothesis that an incipient black hole produced by rapid accretion
onto a neutron star could emit jets of the requisite mean photon energy,
total energy, and collimation.  In the susy phase
transition model it is possible even for an isolated star to collapse
and one avoids the possible problems of how to
rapidly mix the accreted material to kindle fusion and of how
to rapidly extract sufficient energy at the proper wavelengths.

For our chosen stellar mass of $1.09 M_0$ we find
that core collapse happens when $f_e$ drops below $0.88$ at a radius of
$2.7\cdot 10^{-4} R_E$.  The mass at that radius is roughly $0.6 M_0$
, well below
the Chandrasekhar limit, ($1.4 M_0$), below which no black holes would be
expected in the standard model.

\begin{figure}%[htbp]
\begin{center}
\epsfxsize= 4in %% 6.7in % actual
% 4.5in
\leavevmode
\epsfbox{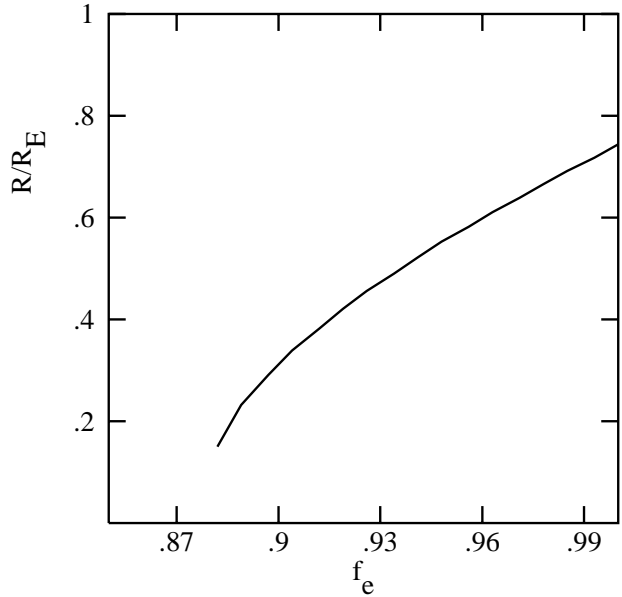}
\end{center}
\caption{The stellar radius in units of $R_E$ as a function of
electron fraction in the toy model of section II.}
\label{stellarradius}
\end{figure}

In figure \ref{stellarradius}, we show the radius of the stable
configuration of
fixed mass as the electron ratio falls below unity down to $0.88$.
One sees that, at this point of core collapse, there
is still a significant electron degeneracy.

\begin{figure}%[htbp]
\begin{center}
\epsfxsize= 4in %% 6.7in % actual
% 4.5in
\leavevmode
\epsfbox{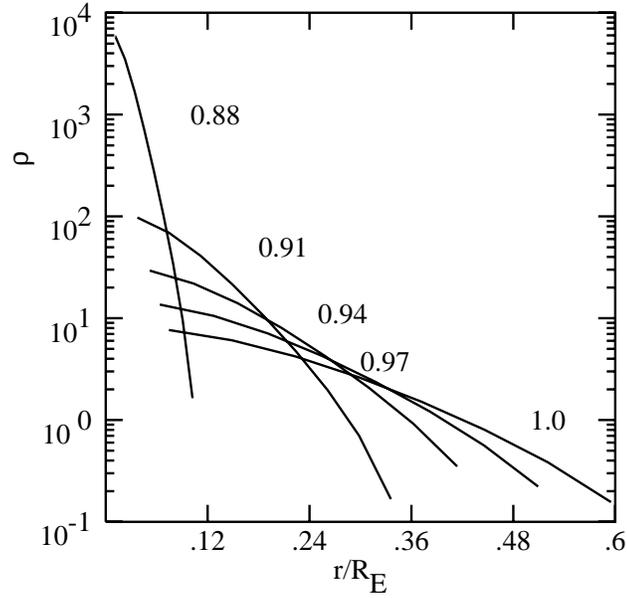}
\end{center}
\caption{The stellar density profile in units of $M_0 R_E^{-3}$
as a function of radius for
five different indicated values of electron fraction $f_e$.
The sharp increase in the central density as $f_e$ approaches $0.88$
is apparent.}
\label{densityprofile}
\end{figure}

Figure \ref{densityprofile} plots density in units of $M_0/R_E^3$ versus
radius in units of $R_E$ for five different values of $f_e$.

\begin{figure}%[htbp]
\begin{center}
\epsfxsize= 4in %% 6.7in % actual
% 4.5in
\leavevmode
\epsfbox{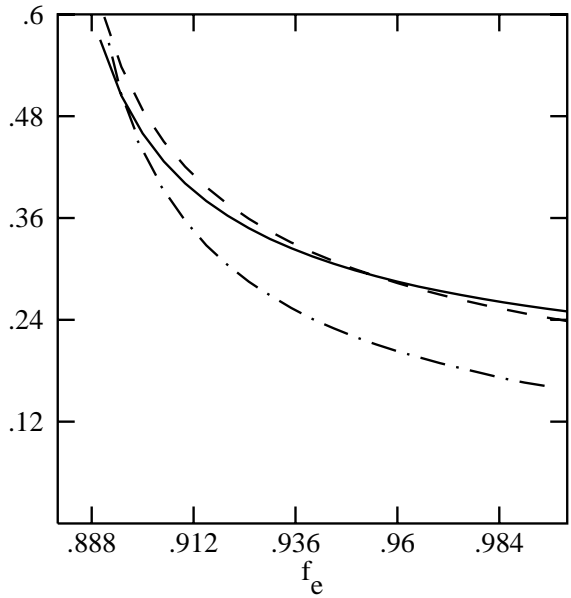}
\end{center}
\caption{various stellar properties as a function of electron fraction.
See text.}
\label{d2feb}
\end{figure}

Figure \ref{d2feb} shows as a function of $f_e$ the total kinetic energy
of the degenerate electron gas in units of $10^{51}$ ergs (solid curve),
the mean electron kinetic energy in MeV (dashed curve), and the
negative of the gravitational energy in units of $10^{55}$ ergs
(dot-dashed curve).  Each
point on the curves represents a zero temperature equilibrium stage in
which the degeneracy pressure gradient balances the gravitational
pressure gradient.  The release of gravitational energy as $f_e$
decreases is more than sufficient to replenish the energy of the
Fermi sea.
From the dot-dashed curve of figure \ref{d2feb} one can see that
about $10^{54}$ ergs of gravitational energy is released before the
point of instability is reached.  How much of this escapes and how
much is swallowed by the incipient black hole is dependent on the
temporal behavior of the collapse which is left to a later analysis.
However, it is clear that the total
radiation is potentially more energetic than the estimates of
\cite{CK},\cite{growth}.  The photons of energy below the Fermi energy
should freely escape from the star.  We can see from the dashed
curve of figure \ref{d2feb}
that the average energy of these photons will typically lie in the
$0.1$ MeV to $1$ MeV as in the observed gamma ray bursts. 
In this section we have neglected the additional energy release from
fusion induced by the gravitational energy dumping.

\section{\bf Growth of the susy phase from a small bubble}
\setcounter{equation}{0}

\begin{figure}%[htbp]
\begin{center}
\epsfxsize= 4in %% 6.7in % actual
% 4.5in
\leavevmode
\epsfbox{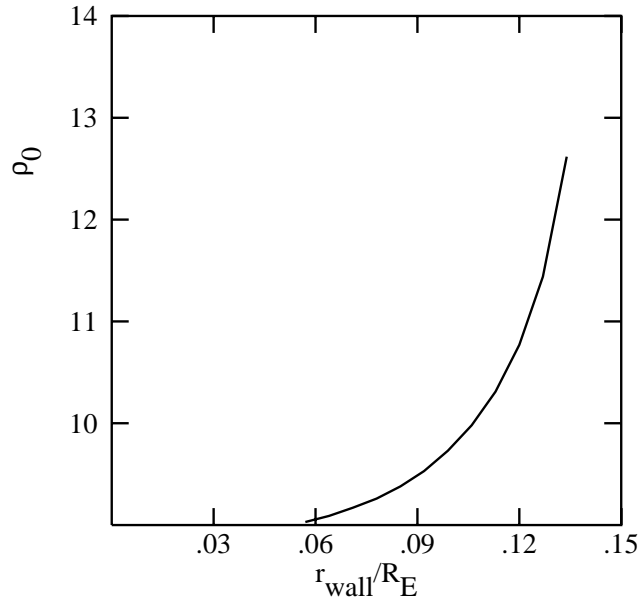}
\end{center}
\caption{The central density in units of $M_0 R_E^{-3}$ as a 
function of bubble wall radius requiring
that the total mass is fixed at $1.09 M_0$.}
\label{centraldensity}
\end{figure}

     In this section we will assume the phase transition takes place
in a small bubble in which the $f_e$ ratio drops immediately to
approximately zero.  The energy release from the Pauli towers in
a Carbon or Oxygen nucleus is estimated to be about $0.5$ MeV per
nucleon.
\be
   \varepsilon = \frac{0.5 {\displaystyle MeV}}{m_N c^2} \approx 5\cdot 10^{-4}
 \quad .
\ee

The radiation pressure is 1/3 of the radiative energy density so the 
radiative pressure gradient is then
\be
    \frac{P_{\it rad}}{dr} = \frac{\varepsilon c^2}{3} \frac{d\rho}{dr}
\ee
Within the susy bubble this radiative pressure gradient must balance
the gravitational pressure gradient eq.\ref{gravgrad}.  In a time
dependent treatment the radiative pressure will gradually decrease and
there will be a gravitational collapse.  In this article we do not
attempt to treat this cooling but leave the time dependent processes to
future study.

    Outside of the bubble we will have $f_e = 1$ and there can be a
balance between the degeneracy and gravitational pressure gradients. 
We consider a sequence of increasing wall radius with $f_e=0$ for 
$r<r_{\it wall}$ and $f_e=1$ for $r>r_{\it wall}$.

In figure \ref{centraldensity} we show the central density as a function
of wall radius for a stellar mass of $1.09$ M$_0$.  As the bubble grows
the central density peaks and, for this mass,
the system becomes unstable at a wall radius of $0.14$ R$_E$ at which no
equilibrium configuration is found.

\begin{figure}%[htbp]
\begin{center}
\epsfxsize= 4in %% 6.7in % actual
% 4.5in
\leavevmode
\epsfbox{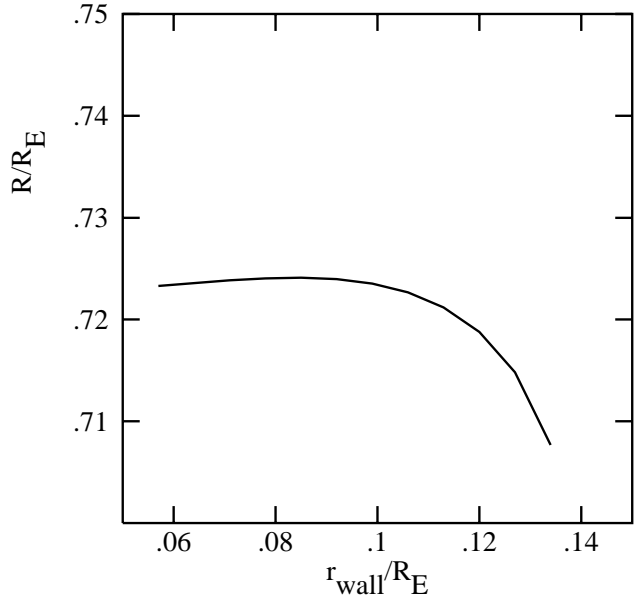}
\end{center}
\caption{The total stellar radius
as a function of wall radius both given in units of earth radius.}
\label{radius}
\end{figure}

As the susy bubble grows the total radius of the star in this model
varies only slowly at first
as shown in figure \ref{radius}.  However, as one approaches the
instability the radius drops sharply.  As discussed in \cite{CK} and
\cite{growth}, the bubble is confined within the star 
and does not grow to engulf the universe.

\begin{figure}%[htbp]
\begin{center}
\epsfxsize= 4in %% 6.7in % actual
% 4.5in
\leavevmode
\epsfbox{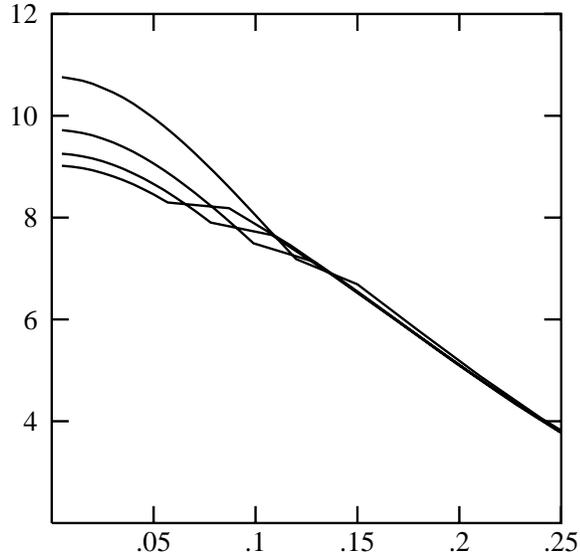}
\end{center}
\caption{The density distribution in units of $M_0 R^{-3}$ as a function of
radius for four values of the wall radius (see text).  The slope of the
density distribution is discontinuous at the bubble wall. }
\label{densitydist}
\end{figure}

As would be expected in the model where the electron fraction $f_e$
is discontinuous at a sharp wall boundary, the equilibrium density
distribution shows a break at the boundary.  In figure \ref{densitydist}
we show the variation of density with radius for four values of 
$r_{\it wall}$, namely $r_{\it wall}$ equal to $0.057 R_E$, $0.078 R_E$,
$0.099 R_E$, and $0.120 R_E$.  The central density increases with wall
radius.

\section{\bf Summary}
\setcounter{equation}{0}

    The analysis of the current paper presents several new aspects
of the behavior of a dense star following a phase transition to exact susy.
Although we have not analysed the time dependence of the problem, the
picture emerging from the present study is as follows.  The conversion
of Fermions to Bosons tends to lift the degeneracy
pressure which supports dense stars in the broken-susy world.  This
tends to decrease the stellar radius and disproportionately increase the
the central density which in turn tends to maintain or increase the mean
and maximum energy of the Fermi gas.  The additional energy is
provided by the shift in gravitational potential energy toward
more negative values.  The gravitional energy released goes first into
replenishing
the Fermi sea.  Additional photons with energy below the Fermi energy
pass freely out of the star while more energetic photons increase the
temperature but are trapped in
regions of high density and only slowly escape.  The consequent radiation
pressure slows the gravitational collapse and increases the amount of
energy that can escape before the star is engulfed in a black hole.
    The Bose enhancement of the selectron final state of the conversion
process has been shown \cite{CP} to lead to some amount of jet structure
but, without replenishment of the Fermi sea, this would not be as efficient
as necessary if the bursts are totally jet-like as suggested by
some.  The release of gravitational energy serves as a natural pumping
mechanism to provide this replenishment.  On the other hand much additional
energy could be released which may not be totally jet-like
but might be part of a quasi-isotropic burst much more
energetic than implied by a purely jet like emission.
    The two simple quasi-adiabatic models presented here suggest that the
Fermi sea is only partially depleted before the star becomes unstable
and collapses.  This supports the idea that the Landau-Pomeranchuk
effect plays an important role in the rapid gamma ray extraction.
    The mean photon energy expected in this model is
in the range of currently observed gamma ray bursts
and the total energy
in the Fermi sea is roughly the observed burst energy assuming a
strong collimation.  The possibility of continually replenishing the
Fermi sea as the susy conversion proceeds allows for the possibility
of emitting more energy than could be obtained by a single
emptying of the Fermi sea.

    The main remaining problem left for future study in the susy star model
is a complete monte-carlo
of the time dependence of the coupled bubble growth and stellar collapse
following the phase
transition.  This should allow a more complete prediction of the total
energy released and resolve the question of jet structure in this model.

    This work was supported in part by the US Department of Energy
    under grant DE-FG02-96ER-40967.

\end{document}